\documentclass[
10pt, twocolumn, prl, superscriptaddress
]{revtex4-1}
\usepackage{graphicx, hyperref, color, bm, amsmath, amssymb, lipsum}  
\usepackage[dvipsnames]{xcolor}
\usepackage[normalem]{ulem}

\newcommand{\BE}[0]{\begin{equation}}
\newcommand{\EE}[0]{\end{equation}}
\newcommand{\BEA}[0]{\begin{eqnarray}}
\newcommand{\EEA}[0]{\end{eqnarray}}
\newcommand{\jkv}[0]{J_K} 
\newcommand{\jhc}[0]{J_H} 
\newcommand{\jhr}[0]{J_{H}'} 
\newcommand{\jkx}[0]{J_{x}} 
\newcommand{\jky}[0]{J_{y}} 
\newcommand{\jkz}[0]{J_{z}} 
\newcommand{\jkxyz}[0]{J_{x,y,z}} 
\newcommand{\enc}[0]{j^{\epsilon}} 
\newcommand{\pol}[0]{P^{\epsilon}} 
\newcommand{\drd}[0]{D}  
\newcommand{\co}[0]{C_0}  
\newcommand{\kdc}[0]{\kappa_{dc}}  
\newcommand{\SR}[0]{\Theta}  
\begin{document}
\title{Spin liquid fingerprints in the thermal transport of a Kitaev-Heisenberg 
   ladder} 
\author{Alexandros Metavitsiadis}\email{a.metavitsiadis@tu-bs.de}
\affiliation{Institute for Theoretical Physics, Technical University 
  Braunschweig, D-38106 Braunschweig, Germany} 
\author{Christina Psaroudaki}\email{christina.psaroudaki@unibas.ch}
\affiliation{Department of Physics, University of Basel, Klingelbergstrasse 82,
  4056 Basel, Switzerland}
\author{Wolfram Brenig}\email{w.brenig@tu-bs.de} 
\affiliation{Institute for Theoretical Physics, Technical University 
  Braunschweig, D-38106 Braunschweig, Germany} 
\date{\today}

\begin{abstract}
We identify fingerprints of a proximate quantum spin-liquid (QSL), observable by
finite-temperature dynamical thermal transport within a minimal version of the
idealized Kitaev model on a two-leg ladder, if subjected to inevitably present
Heisenberg couplings. Using exact diagonalization and quantum typicality, we
uncover (i) an insulator-conductor crossover induced by fracton recombination at
infinitesimal Heisenberg coupling, (ii) low- and high-energy signatures of 
fractons, which survive far off the pure QSL point, and (iii) a non-monotonous current
life-time versus Heisenberg couplings. Guided by perturbation theory, we find
(iv) a Kitaev-exchange induced ``one-magnon'' contribution to the dynamical heat
transport in the strong Heisenberg rung limit.
\end{abstract}

\maketitle

A quantum spin liquid (QSL) is an elusive state of magnetic matter
with the intriguing property of lacking a local magnetic order parameter in the
absence of external fields at any temperature $T$ \cite{0034-4885-80-1-016502, 
RevModPhys.89.025003}. Instead, QSLs may show quantum orders, massive
entanglement and exotic fractional elementary excitations, e.g.  spinons
\cite{PhysRevLett.86.1335, PhysRevLett.98.107204, PhysRevLett.119.137205},
Majorana fermions, gauge vortices \cite{PhysRevLett.114.157202,
PhysRevB.96.205109} and alike.  QSLs are a consequence of frustrating
exchange couplings, such that the local magnetic moments cannot simultaneously
satisfy their mutual interactions \cite{1742-6596-529-1-012001}. 
In a seminal paper \cite{Kitaev20062}, Kitaev
introduced an exactly solvable $\mathbb{Z}_2$ QSL-model, where spin-$1/2$
operators reside on the vertices of a honeycomb lattice and are subject
to exchange frustration from Ising interactions of the type $XX$,
$YY$, or $ZZ$ depending on the direction of the bond
\cite{doi:10.1146/annurev-conmatphys-033117-053934}. Early on, it was proposed
that such patterns can be realized in optical lattices
\cite{PhysRevLett.91.090402}, and shortly after also
in Mott-Hubbard insulators with strong spin orbit coupling (SOC)
\cite{PhysRevLett.102.017205, PhysRevLett.105.027204}.
In the quest for materials which host Kitaev physics, several compounds
have surfaced, e.g.~the iridates $\alpha$-$\mathrm{Na_2IrO_3}$ or
$\alpha$-$\mathrm{Li_2IrO_3}$, as well as $\alpha$-$\mathrm{RuCl_3}$. The latter
systems, however, all order magnetically at low temperatures due to additional
interactions \cite{ScientificReports.6.37925, PhysRevLett.108.127204,
NatureCommunications.7.10273, PhysRevB.85.180403}. Recently, 
$\mathrm{H_3LiIr_2O_6}$ has been synthesized, which reportedly shows no
magnetic order at temperatures $\gtrsim 10^{-4}J$, with $J$ the exchange
interaction \cite{Nature.554.341}.

With low-$T$ magnetic ordering being the common obstacle in real materials,
preempting the putative formation of a QSL, it is of tantamount importance to
identify and interpret fingerprints, genuine to a QSL in systems which are
subject to residual interactions, which obscure QSL behavior. For Kitaev
magnets, this is not trivial and largely under debate
\cite{NatureMaterials.15.733, PhysRevLett.114.147201, 
PhysRevB.95.174429, PhysRevLett.120.077203,
NaturePhysics.12.912, NatureCommunications.8.1152}.  In this endeavor, thermal transport has also been
employed. Unlike to other magnetic systems
\cite{PhysRevLett.90.197002, PhysRevB.81.020405}, 
the longitudinal thermal conductivity $\kappa_{xx}$ in
$\alpha$-$\mathrm{RuCl_3}$ is predominantly phononic with, however, a very
unusual behavior \cite{PhysRevB.95.241112, PhysRevLett.120.067202,
PhysRevLett.120.117204, PhysRevLett.118.187203, 2018arXiv180308162H}. 
Whether this is due to remnants of Majorana fermions is not clear. Stronger 
evidence of Kitaev physics might show up in finite external magnetic fields
(not considered here), because the low temperature magnetic order is 
suppressed \cite{PhysRevLett.119.037201} and it could give rise to a quantized 
thermal Hall conductance \cite{2018arXiv180505022K}. 

Theoretically, thermal transport studies in pure Kitaev QSLs have been performed
via quantum Monte Carlo simulations in 2D \cite{PhysRevLett.119.127204} or via
ED in 1D and 2D \cite{PhysRevB.96.041115, PhysRevB.96.205121}.
Moment expansions might also provide high-temperature analytic results for
thermal transport of pure Kitaev QSLs in the future \cite{PhysRevB.97.064406}. 
Thermal transport was also studied in magnetically {\it
ordered} phases of a Kitaev-Heisenberg model using spin wave calculations
\cite{PhysRevB.95.064410}. However, the impact of isotropic Heisenberg exchange
on thermal transport, perturbing a pure Kitaev QSL, is a completely open
issue. Here our work makes a step forward. We study the thermal transport
properties of a Kitaev-Heisenberg ladder (KHL), using exact diagonalization
(ED), dynamical quantum typicality (DQT), and perturbation theory. By tuning the
exchange between the limits of a pure Kitaev ladder (KL) and a
Heisenberg ladder (HL), we provide a comprehensive view on the
transport properties, while crossing over from a $\mathbb{Z}_2$ QSL into a
conventional dimer magnet with gapped triplon excitations. En route, we
emphasize characteristics which serve to identify Kitaev physics even at
moderate Heisenberg couplings.

\begin{figure}[t!]
\begin{center}
\includegraphics[width= 0.95\columnwidth]{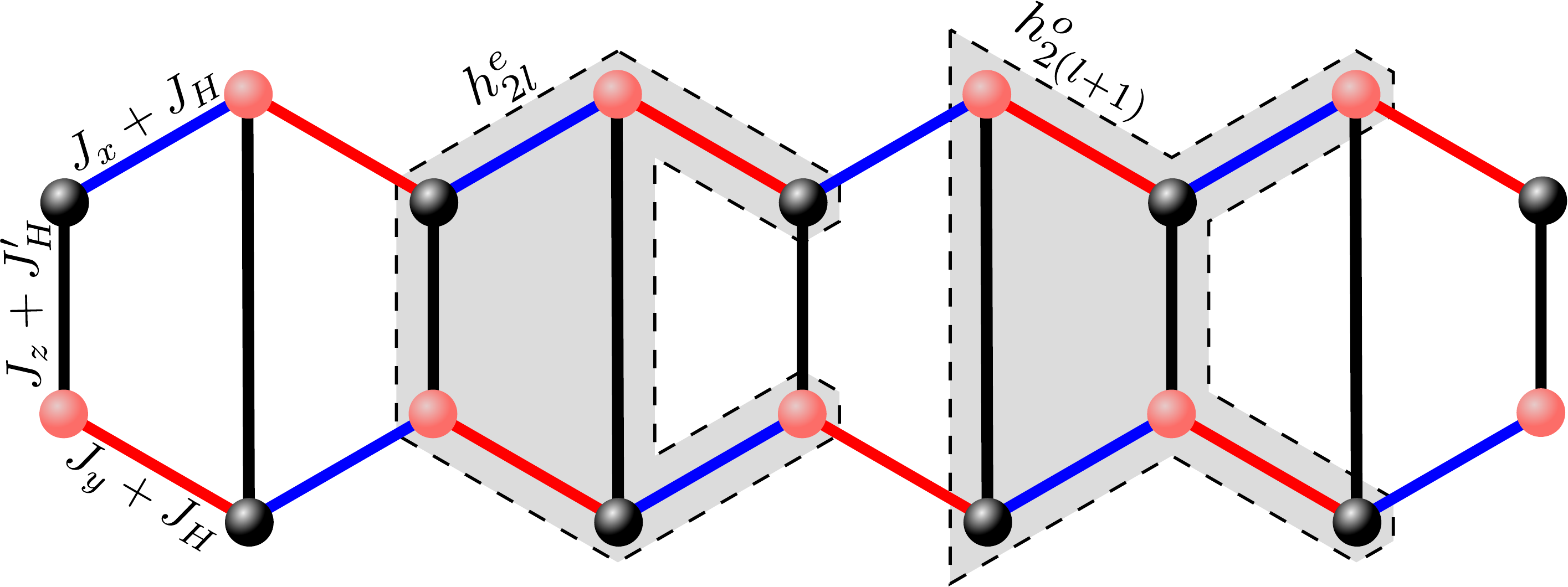} 
\caption{The Kitaev-Heisenberg ladder. $\jkx$, $\jky$, and  $\jkz$  denote 
  Ising interactions, while $\jhc$ and $\jhr$ $\mathrm{SU(2)}$ invariant 
  Heisenberg interactions. The local energy densities $h_l^e$ and  $h_l^o$ 
  used to define the energy current are highlighted.}
\label{fig:Model} 
\end{center}
\end{figure}

The Hamiltonian for the Kitaev-Heisenberg model on a ladder of 
$L$ rungs with boundary conditions  is given by 
\begin{equation} 
  H =  \sum_{a=x,y,z} \sum_{\langle i,j\rangle}    
  J_{ij}^a  S_{i}^a S_{j}^a   
  + J_{ij} S_{i}^a  S_{j}^a~. 
  \label{eq:HS}
\end{equation}
Here $S$ are spin-$1{/}2$ operators, and the restricted sum 
over $i,j$ reproduces the geometry depicted in Fig.~\ref{fig:Model}. 
$J_{ij}^a = \jkx,\jky,\jkz$ denote the anisotropic Kitaev 
interactions--only one of them is non-zero per bond--which we parametrize  
in terms of the coupling strength $\jkv$. On the other hand, 
$J_{ij} = \jhc,\jhr$ are $\mathrm{SU(2)}$ invariant Heisenberg interactions.
If $\jhc\neq\jhr$ is considered we will note this explicitly. 
Lastly, we set the lattice constant equal to unity, as well as the Planck 
and Boltzmann constants. 

In the absence of Heisenberg interactions ($J_{ij}=0$) the system is a 
$\mathbb{Z}_2$ spin liquid \cite{PhysRevLett.98.087204, Wu20123530, 
PhysRevB.96.205109}. The spin degrees of freedom fractionalize into two 
species of Majorana fermions and the Hamiltonian acquires the form \cite{PhysRevB.96.205109} 
\begin{equation}
  H = -\frac{i}{4} \sum_{b} 
  J_x   c_b c_r +  J_y   c_b c_r  
  - J_{z} (i\bar{c}_b \bar{c}_r) c_b c_r \,,
  \label{eq:HF}
\end{equation} 
where $c,\bar c$ represent Majorana fermions,
$\{c_i,c_j\}=2\delta_{ij}=\{\bar{c}_i,\bar{c}_j\}$, while the indices $b$ and
$r$ correspond to the ``black'' and ``red'' sites of the lattice respectively. 
The quantity in the parenthesis is a good quantum number of the model, $\eta =
i\bar{c}_b \bar{c}_r = \pm 1$, and therefore the $\bar c$ species becomes
static. By defining Dirac fermions from pairs of Majorana fermions residing 
on the two sites of the same rung, Hamiltonian \eqref{eq:HF} transforms to a 
tight-binding \emph{chain}  with  pairing  terms in the presence of a
$\mathbb{Z}_2$ gauge field, with the latter acting as a disorder potential. 
The ground state of the system lies in the uniform $\eta$-sectors, 
and it can either be gapless for $|\jkx-\jky| = \jkz$, or gapped otherwise. 
Further we emphasize that for $|\jkx-\jky| > \jkz$ the system acquires a finite 
topological (string-)order parameter \cite{PhysRevLett.98.087204, Wu20123530}.

Transport properties of the KL were analyzed in
Ref.~[\citenum{PhysRevB.96.041115}], showing that in this quasi-1D case, the sole
carriers of heat, the Majorana fermions, scatter from the thermally activated
static gauge disorder such, that localization occurs. I.e.  the KL turns into an
ideal heat insulator at all temperatures. In the pure 2D Kitaev model similar
scattering occurs, but too weak to force localization, leading to normal heat
conduction \cite{PhysRevLett.119.127204, PhysRevB.96.205121}. In contrast,
the HL exhibits a ground state adiabatically connected to a rung-singlet product
(RSP) state, and triplon excitations \cite{Dagotto618}.  
The energy transport of the HL has been analyzed
exhaustively over wide ranges of coupling strengths and temperatures and is well
understood to be diffusive \cite{PhysRevLett.92.067202, PhysRevLett.116.017202}.

To analyze the thermal transport properties of the KHL, we obtain the energy
current operator $\enc$ from the time derivative of the polarization operator,
$\pol = \sum_{l} 2l h_{2l}$ \cite{loc-nla.cat-vn1437043}, which yields $\enc =
-2i\sum_{l} [h_{2l}, h_{2(l-1)}]$. The local energy density $h_{2l}$
satisfies the constraint $\sum_l h_{2l} = H $, rendering the choice of
$h_{2l}$ not unique. In this work, we choose $h_{2l} = (h_{2l}^e+ h_{2l}^o)/2$,
see Fig.~\ref{fig:Model}. We emphasize, that such freedom of choice does not
affect our main findings.

The real part of the energy current correlation function is given by 
$C(t) = \mathrm{Re}[\langle \enc(t)\enc\rangle/L]$, 
where the brackets $\langle \cdots \rangle$ denote the thermal expectation 
value at temperature $T$. The thermal Drude weight (DW) $D$ as well as the 
regular part $\kappa'$ of the thermal conductivity,  
$\kappa(\omega) = 2\pi \drd \delta(\omega) + \kappa'(\omega)$,
are obtained via 
\begin{equation}
  \drd = \frac{\co}{2T^2},~ 
\kappa'(\omega)  = \mathcal{P} \frac{2\beta}{\omega} 
  \tanh \frac{\beta \omega}{2}  
  \int_{0}^\infty dt   \cos\omega t~ C(t), 
  \label{eq:kappa}
\end{equation}
with $\beta=1/T$, $\mathcal{P}$ the principal value,  and $\co$ the 
time independent contribution of degenerate states to $C(t)$. 
The static value of the regular part is determined by the  limiting procedure 
$\kdc =  \kappa'(\omega\rightarrow0)$.
A finite value of $\drd$ signifies dissipationless energy transport, whereas 
the contribution of dissipative modes to the normal dc conductivity is obtained 
by $\kdc$. In the case where $D$ and $\kdc$ vanish 
simultaneously the system is an ideal heat insulator \cite{PhysRevB.53.983}.  

The thermal expectation value is evaluated numerically either using exact 
diagonalization (ED) by tracing over the full Hilbert space, or by using 
the dynamical quantum typicality (DQT), where the full trace is replaced by 
a scalar product of a pure random state $|\psi\rangle$, evolved to 
$|\psi_\beta\rangle = e^{-\beta H / 2}|\psi\rangle $ to account for finite 
temperatures \cite{PhysRevLett.112.120601}. The limiting temperature 
for the DQT is approximately the energy scale of the system $J$, which is 
formally defined below \cite{PhysRevLett.116.017202}. The correlation function 
is then evaluated via $C(t) \approx \mathrm{Re} \frac{\langle\psi_\beta| 
\enc(t) \enc |\psi_\beta\rangle}{L\langle\psi_\beta|\psi_\beta \rangle}$  
by solving a standard differential equation problem for the temperature 
and the time evolution. The error in DQT scales with the inverse square root of 
the Hilbert space dimension, namely exponentially with $L$. 
The time (temperature) evolution is performed with a 
$J\delta t=0.01$ ($J \delta \beta = 0.01$) step [corresponding to an accuracy 
of the order of $O(10^{-8})$ in the fourth order Runge-Kutta algorithm], and 
up to a maximum  time $t_m J = 100\pi$,  giving a $\pi/ t_m= 0.01 J$ frequency 
resolution. We keep the same frequency resolution also for the ED results 
in the  binning of the $\delta$ functions.

\begin{figure}[t!]
\begin{center}
  \includegraphics[width=0.9\columnwidth]{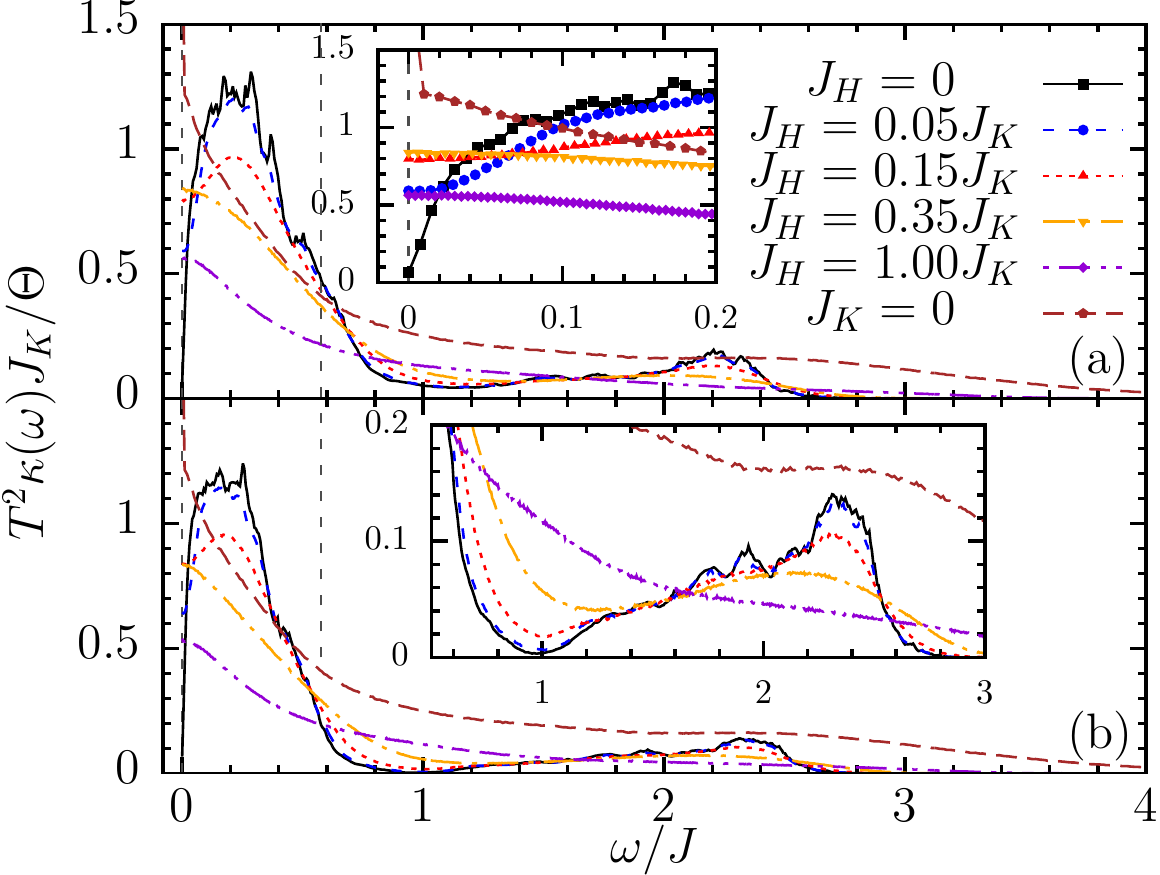} 
\caption{High temperature thermal conductivity versus frequency on a ladder of 
  $L=12$ rungs via DQT for: (a) $\jkxyz=(2,1,1)\jkv$ and (b) 
  $\jkxyz=(3,1,1)\jkv$. For each case the Heisenberg couplings 
  $\jhc/\jkv = 0.05,0.15,0.35,1$ are considered. As a reference, the 
  curves $\jkv=0$ and $\jhc=0$ are also shown, where the latter is obtained 
  using ED in the fermionic representation of Eq.~\eqref{eq:HF}, for a  chain length of 
  $L=32$ sites. The corresponding insets zoom into the low (a) and high frequency (b) 
  parts of $\kappa(\omega)$.
}
\label{figConductorInsulator} 
\end{center}
\end{figure}

Now we detail a central point of this work, i.e. the evolution of the KHL
from the insulating QSL regime of the pure KL to the diffusive one of the HL
close to the RSP state. To this end we present in
Fig.~\ref{figConductorInsulator} the normalized thermal conductivity
$T^2\kappa(\omega)/\SR$, with $\SR=\frac{\pi}{L}\langle \enc \enc \rangle$ the
sum-rule \cite{PhysRevLett.92.067202, PhysRevB.73.085117}. We distinguish two
cases with respect to the pure Kitaev ladder: (a) $\jkxyz = (2,1,1) \jkv$,
corresponding to one of its gapless phases, Fig.~\ref{figConductorInsulator}(a);
(b) $\jkxyz = (3,1,1) \jkv $ corresponding to one of its topological gapped
phases, Fig.~\ref{figConductorInsulator}(b). For each of the two cases, we
present results for $\jhc/\jkv = 0.05,0.15,0.35,1$ derived via DQT at $\beta=0$
on a system of $L=12$ rungs. As a reference, we also present results for the
HL ($\jkv=0$), and for the KL ($\jhc=0$). To reduce large degeneracy effects, 
specific to the latter, we resort to the effective fermionic representation of 
Eq.~\eqref{eq:HF}, in that case, using chains with $L=32$ fermionic sites and 
ED calculations. The frequency axes are rescaled by the ``effective'' coupling
$J=(\jkx+\jky+\jkz)/3+\jhc$.
 
Starting with absent Heisenberg interactions, $\kappa(\omega)$
comprises two prominent structures. First, a low frequency one, which
can be interpreted as the DW, i.e.  the quasiparticle contribution, spread over
a finite frequency region due to the scattering of the itinerant fermions on the
gauge disorder potential. This lifts the degeneracies of the translationally invariant
system yielding a broad low frequency hump. In 1D, itinerant fermions
scattering off a random (here binary) potential leads to insulating behavior
\cite{PhysRev.109.1492}, also for Eq.~\eqref{eq:HF}, 
i.e.~$\drd=0$ and $\kdc= 0$ in the thermodynamic
limit. Consequently, the correlation function exhibits a sharp low frequency dip
and the maximum of $\kappa(\omega)$ is shifted away from $\omega=0$. Second,
a high frequency hump arises due to pair-breaking two-fermion
contributions in $\enc$, which survives at all temperatures--in contrast to the
quasiparticle one which is suppressed at low temperatures due to the fermionic
occupation factors. The two structures are continuously connected in the gapless
case, while in the gapped one the correlation function vanishes for intermediate
frequencies showing that the gap persists even at infinite temperatures.

\begin{figure}[t!]
\begin{center}
  \includegraphics[width=0.95\columnwidth]{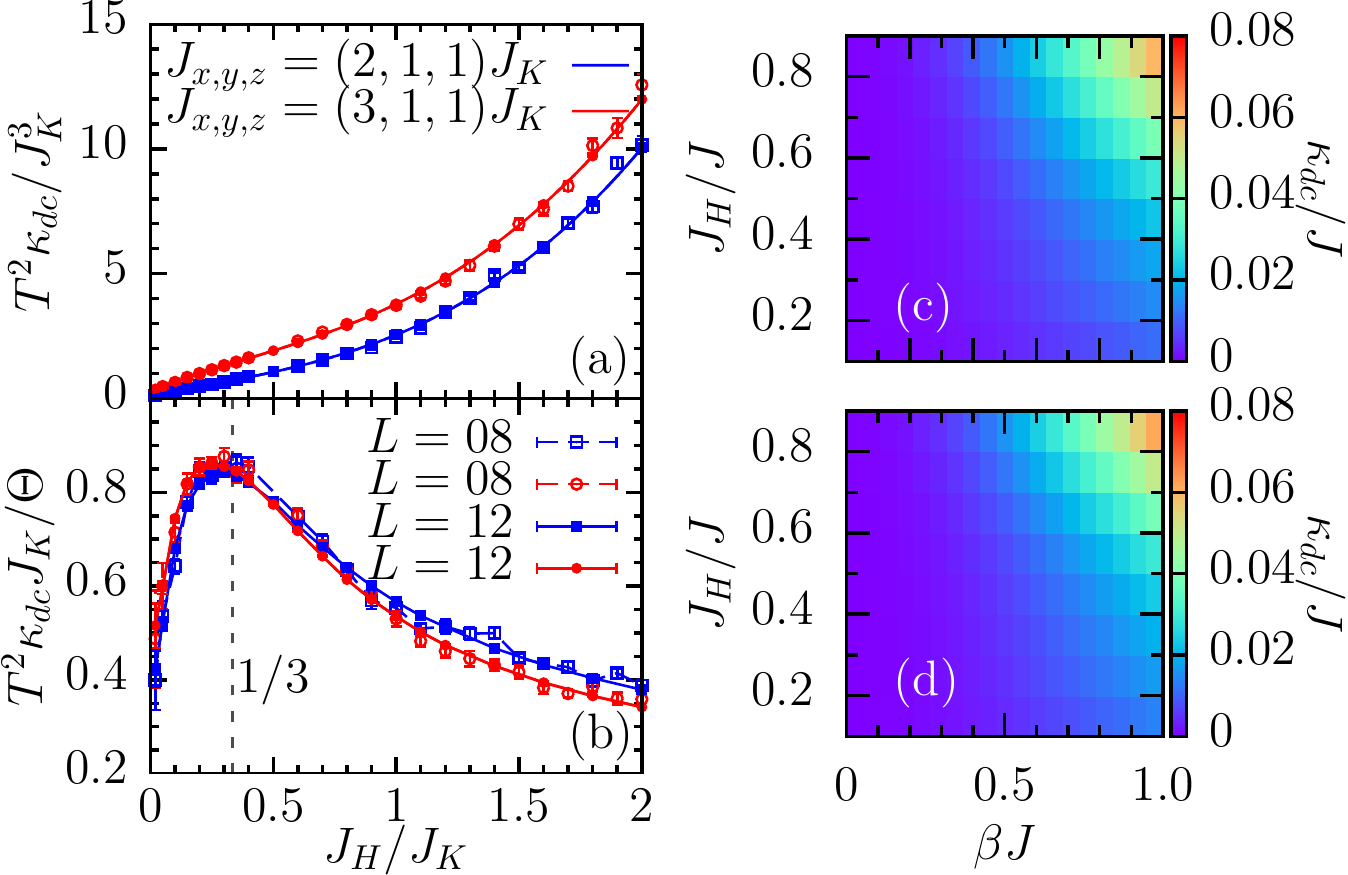} 
  \caption{(a) $T^2\kdc$ and (b) $T^2\kdc/\SR$ 
    versus the ratio $\jhc/\jkv$ at $\beta=0$ via DQT. Open points correspond 
    to $L=8$ while the filled ones to $L=12$. (c) and (d), heatmap of 
    $\kdc$ for $L=10$ versus temperature and the couplings 
    $\jkv$, $\jhc$ keeping $J$ fixed. (c) corresponds to 
    $J_{x,y,z}=(2,1,1)\jkv$ and (d) to $J_{x,y,z}=(3,1,1)\jkv$. 
}
\label{fig:DC} 
\end{center}
\end{figure}

Now we invoke Heisenberg coupling. This breaks the $\mathbb{Z}_2$
symmetry, renders the gauge-fluxes mobile, and restores translational invariance
on some low-energy scale, expanding as $J_H/J_K$ increases. In fact, for all
$J_H\neq 0$ considered, localization breaks down, and a {\it finite} dc
conductivity emerges in Fig.~\ref{figConductorInsulator}(a,b) and the inset of
Fig. \ref{figConductorInsulator}(a). Yet, for a substantial range of
$J_H/J_K\lesssim 0.2$, and on a frequency scale of $O(1)$ the low-$\omega$ hump
and depletion region persists, very suggestive of a fractionalized two component
``liquid'' of light(heavy) mobile Majorana fermions(gauge fluxes). Actually, the
fluxes maintain a finite expectation value for $\jhc\neq0$ \cite{agrapidis}. 
As $\jhc$ is further increased, the system enters the
Heisenberg regime, where the low-$\omega$ depletion is completely filled in
and the correlation function becomes monotonous at low frequencies.
Figs.~\ref{figConductorInsulator}(a,b) nicely support the naive expectation,
that the coupling ratio separating the Kitaev from the Heisenberg regime should
satisfy $\jhc \approx \jkv/3$ even if $\jkx\neq \jky,\jkz$ as in
Figs.~\ref{figConductorInsulator}(b), see also Fig.~\ref{fig:DC}(b).

While the low-$\omega$ depletion-hump structure is   intricately intertwined
with the two-component nature of the fractionalization, the high-$\omega$
pair-breaking peak directly probes only one part of the fractional excitations,
i.e. the two-fermion DOS. As is obvious from the inset of
Fig.~\ref{figConductorInsulator}(b), this feature persists well
into the range of finite Heisenberg interactions, namely $0\leq J_H/J_K \lesssim 0.6$,
thereby providing not only an unequivocal fingerprint of the original KL QSL in
the presence of perturbing Heisenberg exchange, but also a measure for the crossover 
scale $J_H/J_K|_{rec}$, at which Majorana fermions and fluxes recombine to form 
triplons.

Let us now focus on the dc part of the thermal transport, both, versus the
coupling constants, as well as the temperature and for $\jkxyz/\jkv = (2,1,1),
(3,1,1)$, i.e. for a gapless and a gapped case.
We begin with $T^2\kdc$ versus $\jkv/\jhc$ at $\beta=0$ in
Fig. \ref{fig:DC}(a).
A clearly monotonous increase
with increasing $\jhc$ is observable, corroborating not only the insulating
behavior of the KL, but also a {\it critical} coupling for localization of
$J_H=0$. In fact the data can be fitted very well by a fourth order polynomial with
minor offsets, strongly suggesting an insulator as $\jhc\rightarrow 0$. 
Next, we normalize to the sum-rule, displaying $T^2\kdc / \SR$ in
Fig.~\ref{fig:DC}(b). This can be viewed as a rough measure for a zero-frequency
current life-time. Once again, this figure shows a clear scale of $\jhc \approx
\jkv/3$, separating the KL QSL from the HL RSP. The rapid decrease of $\kdc/\Theta$
below this scale is dictated by the onset of localization, i.e. the vanishing of
$\kdc$. This is in sharp contrast to the physics of the HL, where the current
life-time at $\beta = 0$ is a finite constant. Interestingly the two regimes are
connected {\it non-monotonously}. It is tempting to speculate that this may imply a
reduction of current scattering at the crossover to fractionalization.  In passing,
Fig.~\ref{fig:DC}(a),(b) prove that finite size effects are negligible, showing
little difference between $L=8 \mbox{~and~} 12$.

Next, we consider two contour plots of the temperature dependence of $\kdc$
versus $J_H/J$ at $J_{x,y,z}/\jkv=(2,1,1)$ and $(3,1,1)$ in Figs.~\ref{fig:DC}(c) and
(d). The data is represented keeping the effective energy scale $J$ constant. This
figure clearly shows how a {\it low-temperature} regime of enhanced dc conductivity
developing in the upper right hand corner of the plot, as the system recombines
localized Majorana fermions into mobile triplons upon increasing
$J_H/J$.   We note that $\kdc\propto \beta^2$ as
$\beta\rightarrow 0, \forall J_H/J$. This leads to the blue regions in
Figs.~\ref{fig:DC}(c),(d). The main point relating to the latter is, that for $J_H=0$
this region extends over all $\beta$, consistent with an insulator at all
temperatures \cite{PhysRevB.96.041115}. Finally, in view of the similar appearance of
Figs.~\ref{fig:DC}(c),(d), differences between the gapped and gapless case, which are
certainly present for $\beta>1/J$ remain inaccessible to our numerical approach.

\begin{figure}[t!]
\begin{center}
  \includegraphics[width=0.95\columnwidth]{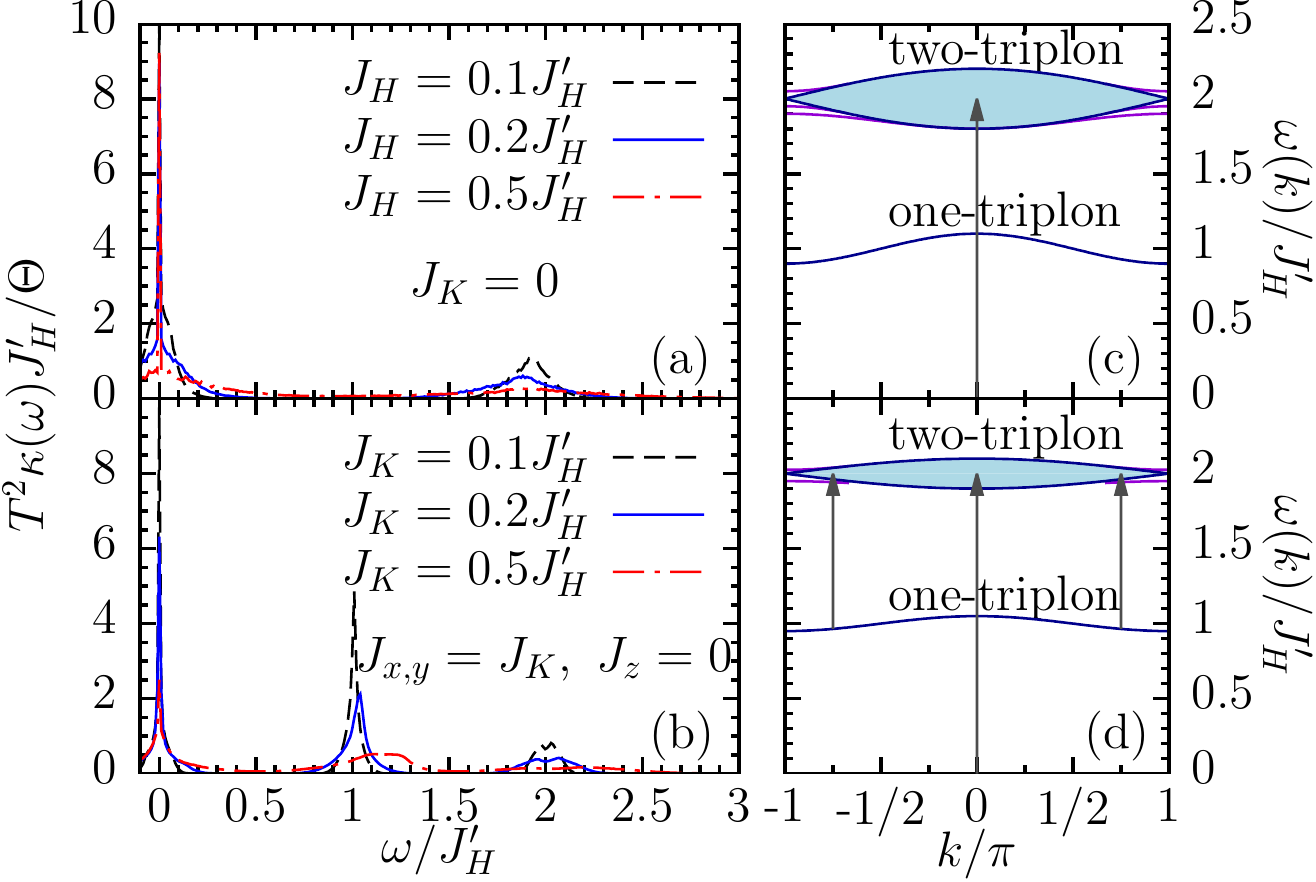} 
  \caption{(a) and (b), $T^2\kappa(\omega)/\SR$ versus frequency in the 
    strong-rung limit for Heisenberg $\jhc/\jhr=0.1,0.2,0.5$ and 
    Kitaev $\jkv/\jhr=0.1,0.2,0.5$ legs respectively obtained via ED at 
    $\beta=0$. (c) and (d), low lying excitation spectrum derived from 
    perturbation theory in the strong-rung limit. The light blue region 
    denotes the two-triplon continuum while the violet dashed lines 
    two-triplon bound states. 
}
\label{fig:SR}  
\end{center}
\end{figure}

Now we change the perspective, and shed light on the impact of Kitaev exchange as
a perturbation, starting from the popular strong rung limit of the HL, i.e. for
$\jhr \gg J_K,\jhc$. For $J_K,\jhc=0$ the ground state $\vert GS \rangle$ is a
RSP state, with energy $E_{GS}/L = -3 \jhr/4$. Finite
$J_K,\jhc$ both shift $E_{GS}$  and induce dispersive
triplon excitations $\vert k,s\rangle$, with momentum $k$ and magnetization $s$.
We evaluate by perturbation theory \cite{0953-8984-2-31-011, 0953-8984-6-43-021, 
PhysRevB.59.6266} the one and two triplon energies, 
\begin{equation}
  \frac{\omega^{(1)}(k)}{\jhr}  = 1 + \lambda \cos k , ~ 
  \frac{\omega^{(2)}(k)}{\jhr}  = 2\left(1 \pm \lambda \cos \frac{k}{2}\right), 
  \label{eq:PT}
\end{equation}
where $\omega^{(2)}(k)$ stands for the boundaries of the two-triplon continuum, 
while $\lambda =\jhc/\jhr$ and $\jkv/2\jhr$, for Heisenberg 
[Fig.~\ref{fig:SR}(c)] and Kitaev [Fig.~\ref{fig:SR}(d)] leg
interactions respectively.  While these 
figures include our results for two-triplon bound states, they will not be considered
further, since they branch off the continuum only near the zone boundary, and
are not expected to contribute significantly to $\kappa(\omega)$
\cite{citeSuppMat}.

From Figs.~\ref{fig:SR}(c),(d) and Eq. (\ref{eq:PT}), we can now interpret ED
for $\kappa(\omega)$ with $L=8$ at $\beta=0$ for Heisenberg, versus Kitaev legs,
in Figs.~\ref{fig:SR}(a) versus (b). In both cases intensity at $\omega\sim 0$
arises from thermally populated triplon states, comprising a Drude weight on
finite systems. Additionally however, for Heisenberg legs $\enc \vert GS
\rangle$ generates a state in the two triplon manifold, which combined with the
selection rules $\Delta S_z=0$ and $\Delta k=0$, dictated by the symmetries of
the Heisenberg Hamiltonian, results in transitions with energies $\omega \sim
2\jhr(1\pm \jhc/\jhr)$. This is clearly seen in Figs. \ref{fig:SR}(a). In sharp
contrast, Kitaev legs induce an additional current mode at $\omega \sim \jhr$,
visible in Fig.~\ref{fig:SR}(b). This qualitative change is a direct consequence
of the loss of $SU(2)$ invariance, which allows for heat-current transitions
between one- and two-triplon states, subject to a $\Delta k=0$ 
only. This one-triplon current intensity will feature a strong temperature
dependence $\sim\exp(-\jhr/T)$ as it involves only excited states. As
Figs. \ref{fig:SR}(a),(b) show our interpretation remains intact up to fairly
strong leg couplings $J_{K,H}/\jhr \approx 0.5$. Finally, we note that $\enc
\vert GS \rangle$ also induces low intensities at three triplon energies, not
considered here.

To summarize,  we have uncovered 
several fingerprint of a proximate Kitaev QSL manifested at various energy
scales in the dynamical thermal transport of the Kitaev-Heisenberg ladder. While
born out of a quasi-1D model study, our results should be transferable to 2D
except for the singular behavior at $\jhc=0$ due to the difference between
localization in 1D and 2D. We hope this may stimulate experiments, realizing
that not only dc thermal conductivity is a well established experimental probe, but
also dynamical heat transport can be addressed, e.g. via fluorescent flash
methods \cite{PhysRevLett.110.147206, OTTER2009796} or pump-probe techniques
\cite{doi:10.1063/1.2937458}. Moreover, a ``tuning'' of the  
exchange couplings, discussed here theoretically, is also experimentally 
feasible - within certain limits - by chemical substitution or external pressure.

\paragraph{Acknowledgments.}
We thank C. Hess, R. Stei\-ni\-ge\-weg, and M. Vojta, for fruitful
discussions. Work of W.B. has been supported in part by the DFG through SFB
1143, proj. A02, and by QUANOMET and CiNNds.  W.B. also acknowledges kind
hospitality of the PSM, Dresden.
%
%
%
%
%
%
\end{document}